\RequirePackage{ifpdf}

\documentclass{PoS_oldRef}

\title{Radio Jets in Young Stellar Objects with the SKA}

\ShortTitle{Radio Jets in YSOs with the SKA}

\author{\speaker{Guillem Anglada}\\ %\thanks{A footnote may follow.}\\
	Instituto de Astrof\'\i sica de Andaluc\'\i a, CSIC, Spain\\
        E-mail: \email{guillem@iaa.es}}

\author{Luis F. Rodr\'\i guez\\
        Centro de Radioastronom\'\i a y Astrof\'\i sica, UNAM, M\'exico\\
        E-mail: \email{l.rodriguez@crya.unam.mx}}

\author{Carlos Carrasco-Gonz\'alez\\
        Centro de Radioastronom\'\i a y Astrof\'\i sica, UNAM, M\'exico\\
        E-mail: \email{c.carrasco@crya.unam.mx}}

\abstract{Jets and outflows are ubiquitous in the process of formation of 
stars since accretion is intimately associated with outflow. Free-free radio 
continuum emission in the centimeter domain is associated with these jets. 
The emission is weak, and sensitive telescopes are required to detect it.

One of the key problems in the study of outflows is to determine how 
they are accelerated and collimated. Observations in the cm range are 
most useful to trace the base of the ionized jets, close to the young 
central object and its accretion disk, where optical or near-IR images 
are obscured by the high extinction present. Radio recombination lines 
in jets (in combination with proper motions) should provide their 3D 
kinematics at very small scale (near their origin). SKA will be crucial 
to perform this kind of observations.

Thermal jets are associated with both low and high mass protostars. The 
ionizing mechanism of these radio jets appears to be related to shocks in 
the associated outflows, as suggested by the observed correlation between 
the centimeter luminosity and the outflow momentum rate. From this 
correlation and that with the bolometric luminosity of the driving star it 
will be possible to discriminate with SKA between unresolved HII regions and 
jets, and to infer physical properties of the embedded objects.

Some jets associated with young stellar objects (YSOs) show indications 
of non-thermal emission (negative spectral indices) in part of their 
lobes. Linearly polarized synchrotron emission has been found in the jet 
of HH 80-81, allowing us to measure the direction and intensity of the 
jet magnetic field, a clue ingredient in determining the collimation and 
ejection mechanisms. As only a fraction of the emission is polarized, 
very sensitive observations such as those that will be feasible with SKA 
are required to perform these studies.

Jets are common in many kinds of astrophysical scenarios. Characterizing 
radio jets in YSOs, where thermal emission allows us to determine their 
physical conditions in a reliable way, would be also useful in understanding 
acceleration and collimation mechanisms in all kinds of astrophysical jets.
 }

\FullConference{
Advancing Astrophysics with the Square Kilometre Array\\
June 8-13, 2014\\
Giardini Naxos, Italy}

\newcommand{\skipthis}[1]{}

\newcommand\apj{ApJ}
\newcommand\apjl{ApJL}
\newcommand\rmxaa{RevMexAA}
\newcommand\aap{A\&A}

\begin{document}

\section{Introduction}

In the last years it has become clear that collimated outflows are 
present in young stars across all the stellar spectrum, from O-type 
protostars to brown dwarfs, suggesting that the disk-jet scenario is 
valid to describe the formation of stars of all masses.

Outflows associated with young stars frequently exhibit a central weak 
centimeter emission source (Anglada et al. 1992; Anglada 1995; Rodr\'\i 
guez \& Reipurth 1998).
 In the best studied cases, these sources are resolved angularly at the 
sub-arcsec scale and found to be elongated in the direction of the 
large-scale tracers of the outflow (e.g. Rodr\'\i guez et al. 1990; 
Rodr\'\i guez 1995, 1996; Anglada 1996), indicating that they trace the 
region, very close to the exciting star, where the outflow phenomenon 
originates. The centimeter flux density usually rises slowly with 
frequency (positive spectral index).
 Given their morphology and spectrum these sources are usually referred 
to as ``thermal'' radio jets. These radio jets have been found during 
all the stages of the star-formation process, from Class 0 protostars to 
stars associated with transitional disks, suggesting that outflow and 
accretion are intimately related (Rodr\'\i guez et al. 2014).

\begin{figure}[h]
 \hspace*{\fill}\includegraphics[width=0.34\textwidth]{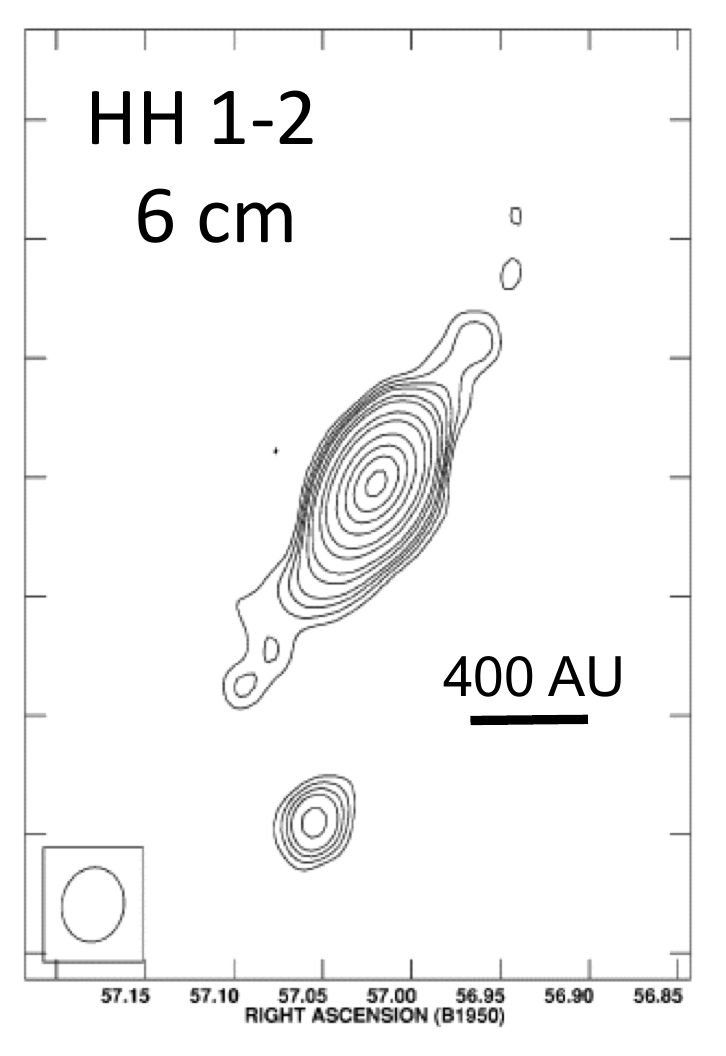}\hfill\includegraphics[width=0.48\textwidth]{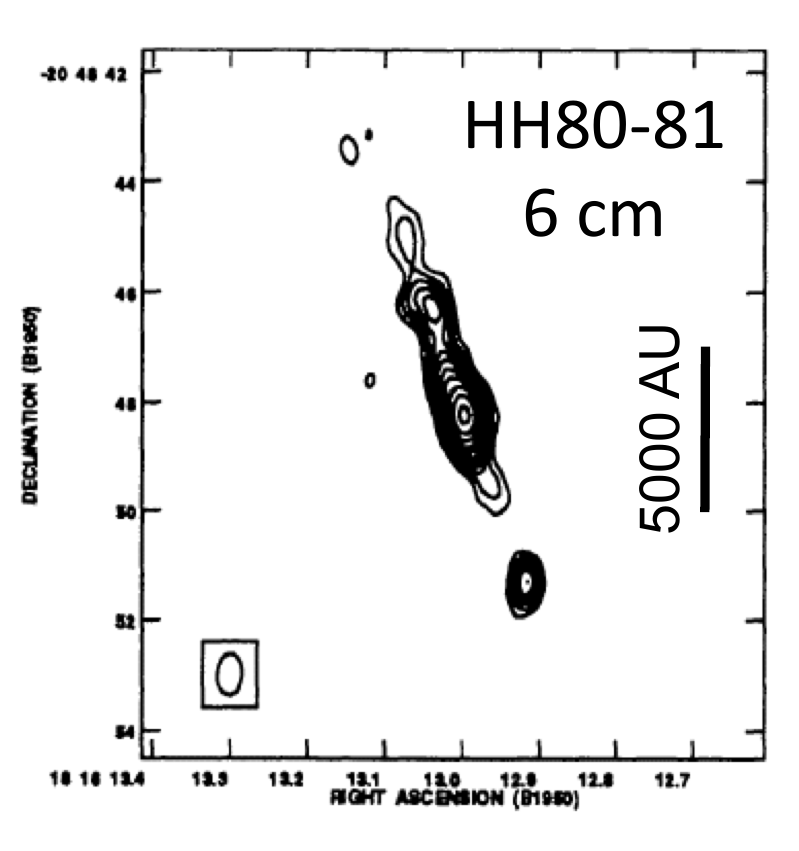}\hspace*{\fill}
 \caption{(Left) HH1-2 radio jet, associated with a low-mass object 
(Rodr\'\i guez et al. 2000). (Right) HH80-81 radio jet, associated with 
a high-mass object (Mart{\'{\i}} et al. 1993).}\label{fig1}
 \end{figure}

With the extraordinary sensitivity of the Jansky VLA, e-MERLIN, and the 
SKA it is expected that all nearby (a few kpc) young stellar objects 
(YSOs) known to be associated with outflows, both molecular and/or 
optical/infrared, will be detectable as centimeter sources (Anglada et 
al. 2015, in preparation). The topic of radio jets from young stars has 
been reviewed by Anglada (1996) and Rodr\'\i guez (1997; 2011).

\section{Properties of Radio Jets}

The study of jets associated with young stars at radio wavelengths is 
important in several aspects. Given the large obscuration present 
towards the very young stars, the detection of the radio jet provides so 
far the best way to obtain their accurate positions. These observations 
also provide information on the direction and collimation of the gas 
ejected by the young system in the last few years (see Table 1) that can 
be compared with the gas in the molecular outflows and optical/infrared 
HH jets, which traces the ejection over timescales several orders of 
magnitude larger. This comparison allows us to make evident the changes 
in the ejection direction, possibly resulting from precession or orbital 
motions in binary systems (Rodr\'\i guez et al. 2008; Masqu\'e et al. 
2012).

The radio jets from YSOs can be described using the free-free models of 
Reynolds (1986). In particular, for a jet of constant temperature, 
velocity and ionization fraction, these models predict that the flux 
density increases with frequency as $S_\nu \propto \nu^\alpha,$ where 
$\alpha = 1.3 - 0.7/\epsilon,$ and $\epsilon$ is the index of the 
power-law dependence of the jet width with distance to its origin. The 
case of $\epsilon=1$ corresponds to a conical (constant opening angle) 
jet. The angular size of the major axis of the jet decreases with 
frequency as $\theta_\nu \propto \nu^{-0.7/\epsilon} = \nu^{\alpha - 
1.3}$. This behavior in flux density and size can be tested with 
multi-frequency observations.

With the Reynolds (1986) models it is also possible to determine from 
the observations the ionized mass loss rate, $\dot M_i$, in the jet and 
the radius at which ionization starts, $r_0$. The jet velocity, $V_j$, 
necessary to determine $\dot M_i$ can be estimated from proper motion 
measurements obtained from multi-epoch observations. In Table 1 we 
present the parameters of a few selected radio jets. These well-known 
jets have centimeter flux densities of 1 mJy or more. However, most of 
the jets associated with low mass stars or even brown dwarfs present 
centimeter flux densities in the order of tens of $\mu$Jy (Rodr\'\i guez 
et al. 2015, in preparation) and their detection and study will require 
of the high sensitivity of SKA.

\begin{table}[h]
{\footnotesize
\caption{Properties of Selected Angularly Resolved Radio Jets in YSOs}
\label{tab1}
\begin{tabular} %{lccccccccccccl}
{l@{\extracolsep{0.3em}}c@{\extracolsep{0.1em}}c@{\extracolsep{0.4em}}c@{\extracolsep{0.4em}}c@{\extracolsep{0.4em}}c@{\extracolsep{0.4em}}c@{\extracolsep{0.4em}}c@{\extracolsep{0.4em}}c@{\extracolsep{0.4em}}c@{\extracolsep{0.4em}}
c@{\extracolsep{0.4em}}c@{\extracolsep{0.4em}}c@{\extracolsep{0.4em}}l}
\hline\noalign{\smallskip}
 {Source}
&  {$L_{\rm bol}$}
&  {$M_\star$}
&  {$d$}
&  {$S_\nu$}
&   
& {$\theta_0$}
& {Size}
& {$V_j$}
& {$t_{\rm dyn}$}
& 
& {$\dot M_i$} 
& {$r_0$} 
& {}\\
& {($L_\odot$)}
& {($M_\odot$)}
& {(kpc)}
& {(mJy)}
& {$\alpha$}
& {(deg)}
& {(AU)}
& {(km s$^{-1}$)}
& {(yr)} 
& {$\epsilon$}
& {($M_\odot$ yr$^{-1}$)}
& {(AU)}
& {Refs.} \\
\noalign{\smallskip}\hline\noalign{\smallskip}
HH 1-2 VLA1 & 20 & $\sim$1 & 0.4 & 1 & 0.3 
& 19& 200& 270 & 2 & 0.7 & $1 \times 10^{-8}$ & $\leq$11 & 1, 2, 3, 4 \\
NGC 2071-IRS3 & $\sim$500 & 4 & 0.4 & 3 & 0.6
&40& 200& 400$^a$ & 3 & 1 & $2 \times 10^{-7}$  & $\leq$18 & 5, 6, 2, 7 \\
Cep A HW2 & 1$\times$$10^4$ & 15 & 0.7 & 10 & 0.7  
&14& 400& 460 & 3 & 0.9 &$5 \times 10^{-7}$ & $\leq$60 & 8, 9, 10, 11, 12 \\
HH 80-81 & 2$\times$$10^4$ & 15 & 1.7 & 5 & 0.2
&34& 1500& 1000 & 7 & 0.6 &  $1 \times 10^{-6}$ & $\leq$25 & 13,\,14,\,15,\,16,\,17,\,18\\
\noalign{\smallskip}\hline
\end{tabular}

$^{a}${Assumed.} \\
 References: {(1) Fischer et al. 2010; (2) Menten et al. 2007; (3) 
Rodr\'{\i}guez et al.\ 1990; (4) Rodr\'{\i}guez et al.\ 2000; (5) Butner et 
al. 1990; (6) Carrasco-Gonz\'alez et al. 2012a; (7) Torrelles et al. 1998; 
(8) Hughes et al. 1995; (9) Patel et al. 2005; (10) Dzib et al. 2011; (11) 
Curiel et al. 2006; (12) Rodr\'{\i}guez et al.\ 1994; (13) Aspin \& Geballe 
1992; (14) Fern\'andez-L\'opez et al. 2011; (15) Rodr\'{\i}guez et al.\ 
1980; (16) Mart\'{\i} et al. 1995; (17) Mart\'{\i} et al. 1993; (18) 
Carrasco-Gonz\'alez et al. 2012b.}
 }
\end{table}

Sensitive, very high angular resolution observations at centimeter 
wavelengths can trace the base of the jets down to the injection radius, 
at scales of a few AUs, where the ionized jet is expected to begin. 
Exploration of this inner region will eventually shed new light on the 
jet acceleration and collimation mechanisms, helping to distinguish 
between different theoretical approaches such as the X-wind (Shu et al. 
2000) and disk-wind (K\"onigl \& Pudritz 2000) models. Shang et al. 
(2004) use the X-wind model to calculate the expected properties of the 
centimeter free-free emission of YSO jets. A comparison of detailed 
modeling results and high quality observations can provide an accurate 
description of the jet physical parameters.

Additionally, we note that a good knowledge of the jet properties is 
indispensable for grain growth studies in protoplanetary disks to 
separate the dust emission of the disk from the free-free emission of 
the jet. It is known that even transitional disks present free-free 
emission in their central regions (Rodr\'\i guez et al. 2014).

\section{Bolometric Luminosity, Outflow, and Radio Continuum Correlations}

Photoionization does not appear to be the ionizing mechanism of radio jets, 
since in the sources associated with low-luminosity objects, the number of 
UV photons from the star is clearly insufficient to produce the ionization 
required to explain the observed radio continuum emission (e.g., Rodr\'\i 
guez et al. 1989; Anglada 1995). The radio luminosity of radio jets ($S_\nu 
d^2$) is correlated with the bolometric luminosity of the source, $L_{\rm 
bol}$, and with the momentum rate in the molecular outflow, $\dot P$ 
(Anglada 1995, 1996; see Fig. 2):

\begin{equation} {\biggl(\frac{S_\nu d^2}{\rm mJy~kpc^2}\biggr)} = 0.008 
\biggl(\frac{L_{\rm bol}}{L_\odot}\biggr)^{0.6}, \end{equation}

\begin{equation} {\biggl(\frac{S_\nu d^2}{\rm mJy~kpc^2}\biggr)} = 190 
\biggl(\frac{\dot P}{M_\odot~{\rm
yr^{-1}~km~s^{-1}}}\biggr)^{0.9}. \end{equation}

\begin{figure}[h]
\includegraphics[width=.5\textwidth]{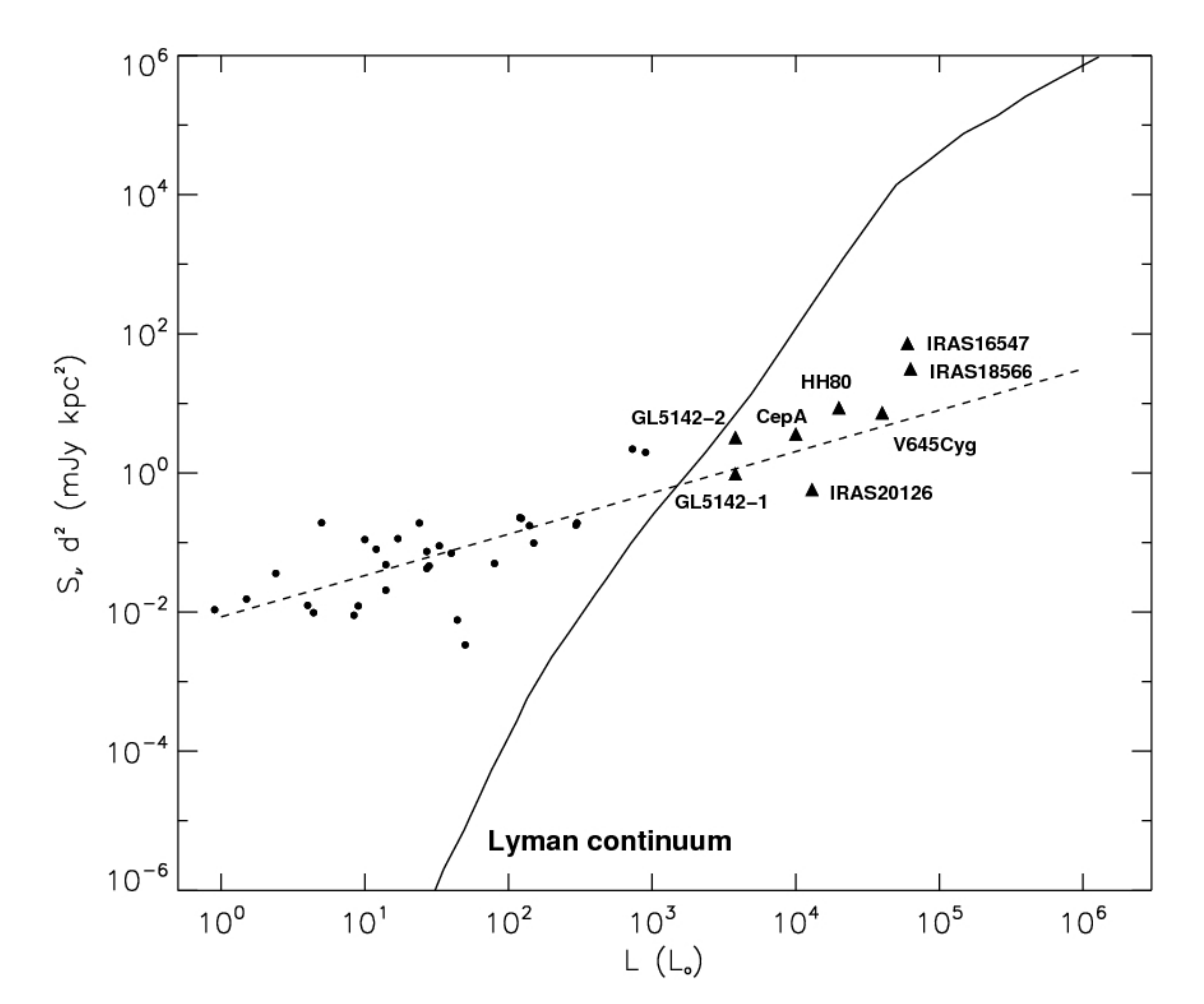}\includegraphics[width=.5\textwidth]{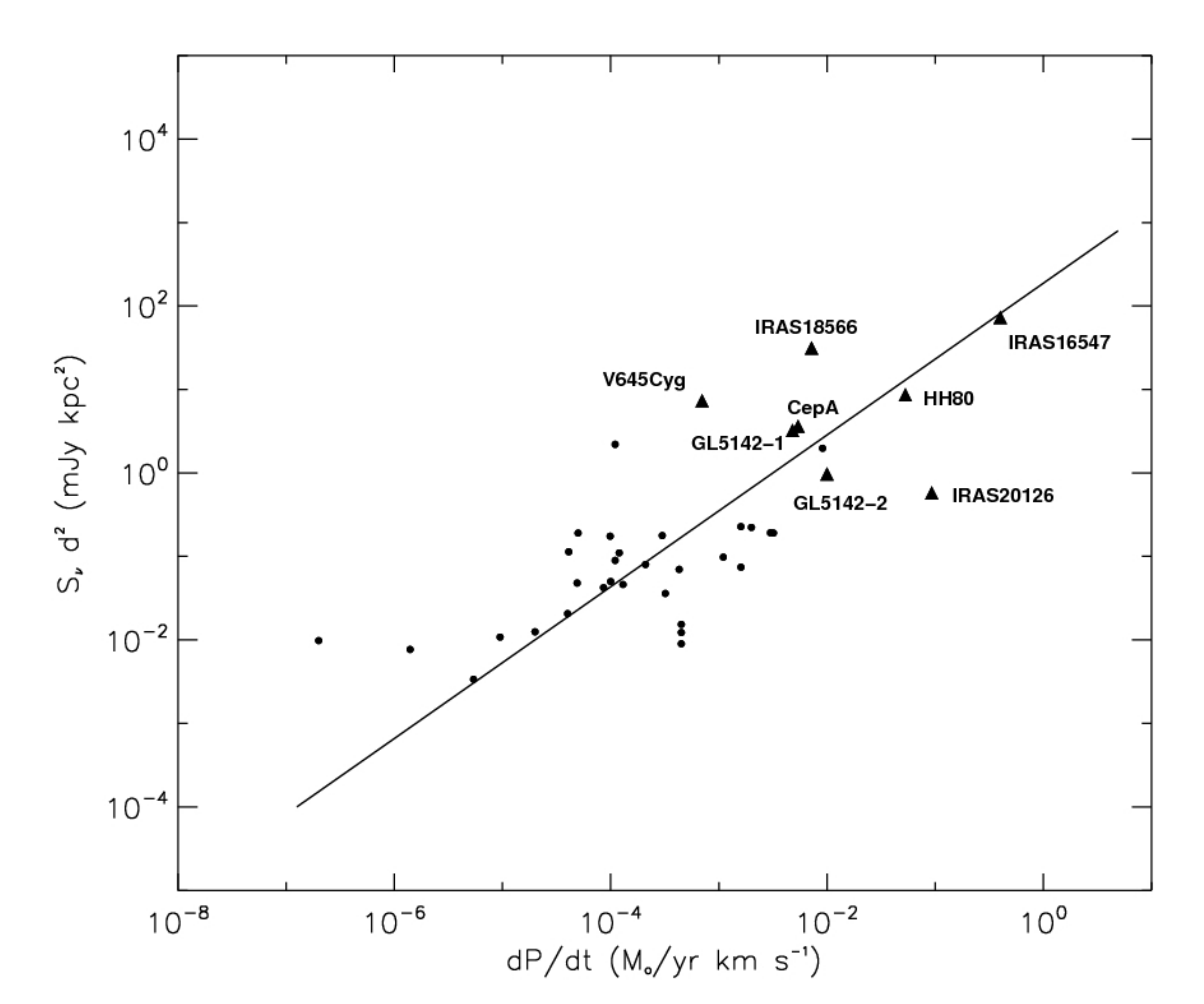}
 \caption{(Left) Radio luminosity versus bolometric luminosity correlation 
for young stellar objects (dashed line). The solid line (Lyman continuum) 
represents the radio luminosity expected from photoionization. (Right) Radio 
luminosity versus outflow momentum rate (solid line). The high luminosity 
objects are labeled in the figures.}\label{fig2}
 \end{figure}

These correlations include young stellar objects with luminosities 
spanning from 1 to $10^5$ $L_\odot$. In particular, these diagrams can 
be used to discriminate between the thermal radio jets (that should 
follow these correlations) and HII regions (that should fall close to 
the Lyman continuum line). Also, detection of radio recombination lines 
(RRLs, see Sect. 4) in jets could be useful to distinguish between jets 
and HII regions, since lines are expected to be broader in jets, as 
noted by Hoare et al. (2007) (see their Fig. 6).

\section{Radio Recombination Lines}

Radio jets are expected to show RRLs as part of the emission processes 
of the plasma. RRLs in jets (in combination with proper motions) should 
provide their 3D kinematics at very small scale (near their origin).

Assuming local thermodynamic equilibrium (LTE) for line and continuum and a 
standard biconical jet with constant velocity and ionized fraction, the 
expected line to continuum flux density ratio in the centimeter wavelength 
regime is given by (Rodr\'\i guez et al. 2015, in preparation):

\begin{equation}
{{S_L} \over {S_C}} = 0.19 \biggl({{\nu_L} \over {\rm GHz}} \biggr)^{1.1}
\biggl({{T} \over {\rm 10^4~K}} \biggr)^{-1.1} \biggl({{\Delta V} 
\over {\rm km~s^{-1}}}\biggr)^{-1} (1 + Y^+)^{-1}. 
\end{equation}

In the above equation, $\nu_L$ is the frequency of the line, $\Delta V$ 
the full width at half maximum of the line, and $Y^+$ is the ionized 
helium to ionized hydrogen ratio. Since the lines from a jet are 
expected to be wide, $\Delta V \simeq$ 100 km s$^{-1}$, their 
detectability will be challenging. For example, at 10 GHz the 
line-to-continuum ratio is expected to be $\sim$0.02. Thus, for a jet 
with a continuum flux density of 1 mJy, the peak line flux density will 
be only $\sim$20 $\mu$Jy. So far, there are no detections of RRLs in 
jets at the expected LTE level. Jim\'enez-Serra et al. (2011) report the 
detection, at millimeter wavelengths, of broad recombination maser lines 
toward the jet in Cepheus A HW2 with flux densities about 5 times larger 
than those expected for LTE. Additional observations are needed to 
understand the nature of RRLs from jets and start using them as tools to 
study the outflow kinematics.
     
\section{Non-thermal Emission}

As discussed above, jets from YSOs have been long studied at radio 
wavelengths through their thermal free-free emission which traces the 
base of the ionized jet, and shows a characteristic positive spectral 
index. However, in the last two decades, negative spectral indices have 
been found in some regions of YSO jets (e.g., Rodr\'\i guez et al. 1989, 
2005; Curiel et al. 1993; Mart\'{\i} et al. 1993; Garay et al. 1996; 
Wilner et al. 1999; see Fig. 3 left). This negative spectral-index 
emission is usually found in strong radio knots, that usually appear in 
pairs, sometimes found moving away from the central protostar at 
velocities of several hundreds of kilometers per second. Because of 
these characteristics, it has been proposed that these knots would be 
tracing strong shocks of the jet against dense material in the 
surrounding molecular cloud. Their negative spectral indices have been 
interpreted as indicating non-thermal synchrotron emission from a small 
population of relativistic particles that would be accelerated in the 
ensuing strong shocks.

\begin{figure}[h]
\includegraphics[width=.43\textwidth]{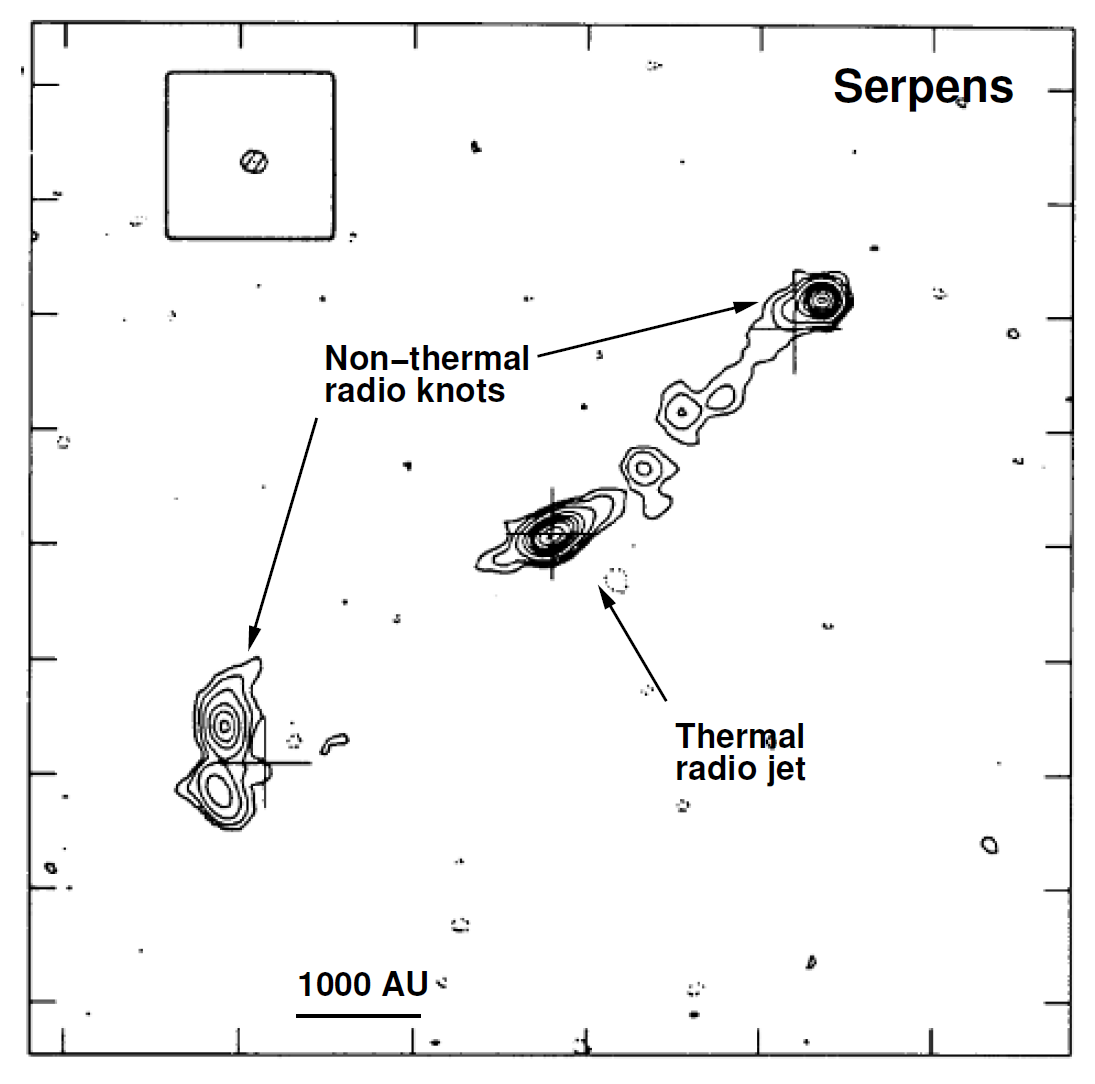}\hfill\includegraphics[width=.52\textwidth]{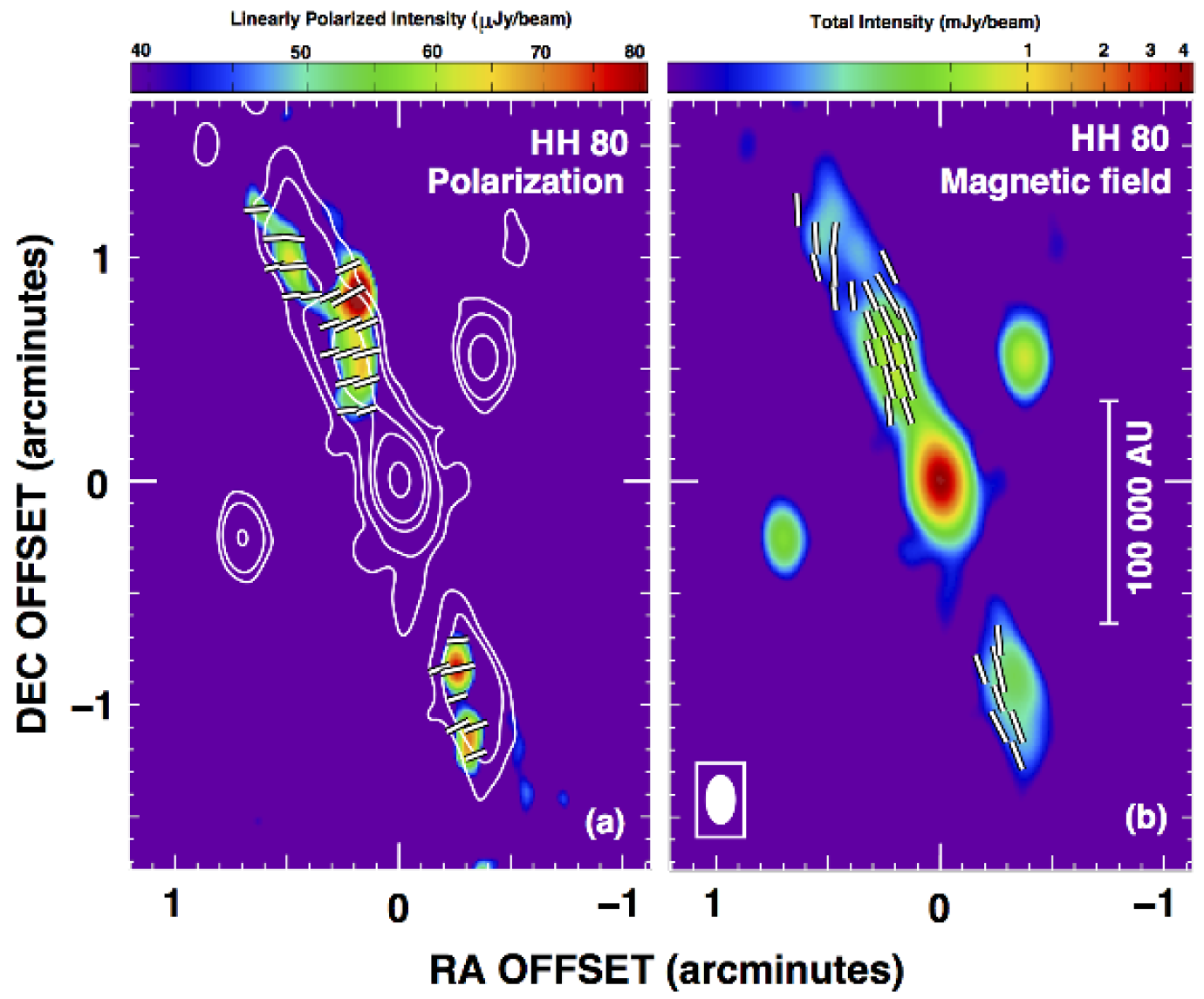}
 \caption{(Left) VLA image at 3.6 cm of the radio jet in Serpens, one of 
the first YSO jets where synchrotron emission was proposed to arise as a 
result of strong shocks with the ambient medium. The central source 
shows a positive spectral index at radio wavelengths, consistent with 
free-free emission, while the two bright knots at the end of the lobes 
show negative spectral indices indicative of thermal emission (adapted 
from Curiel et al. 1993). (Right) VLA images at 6 cm of the HH 80-81 jet 
showing the first detection of linearly polarized synchrotron emission 
in a YSO jet. Panel (a) shows in color scale the polarized emission and 
the polarization direction as white bars. Panel (b) shows the direction 
of the magnetic field as white bars. The total continuum intensity is 
shown in contours (a) and in color scale (b) (from Carrasco-Gonz\'alez 
et al. 2010).}\label{fig3}
 \end{figure}

This scenario has been confirmed only very recently, with detection of 
linearly polarized emission from the HH 80-81 jet (Carrasco-Gonz\'alez 
et al. 2010; see Fig. 3 right). This result provided for the first time 
conclusive evidence for the presence of synchrotron emission in a YSO 
jet making it possible the direct measure and study of the properties of 
the magnetic field strength and morphology. Measuring linear 
polarization in YSO jets is difficult because it is only a fraction of 
the total emission.
 Ultrasensitive radio interferometers, such as the SKA, will allow us to 
detect and image the magnetic field in a large sample of YSO jets. In 
combination with the physical parameters (density, temperature, 
velocity) derived from observations of the thermal component, the 
measurement of the magnetic field from the non-thermal component will 
help in understanding YSO jet acceleration and collimation mechanisms, 
that appear to be similar for all kinds of astrophysical objects.

\section{Expectations for SKA Observations of Radio Jets}

The study of radio jets will greatly benefit from the higher frequency 
bands of SKA. In particular, given the rising spectrum and compact size 
of the radio jets, we anticipate that most observations of these sources 
will be carried out in band 5. SKA will allow us to survey the southern 
hemisphere for radio jets associated with young stars across the mass 
spectrum, including proto-brown dwarfs. At a distance of 500 pc, a radio 
jet from a proto-brown dwarf is estimated to have a flux density of 12 
$\mu$Jy at 10 GHz (Palau et al. 2015, in preparation). The SKA1-MID will 
detect this weak emission at the 10-$\sigma$ level in only 20 minutes of 
on-source integration time. Detection of the jets from the most massive 
protostars (with expected radio luminosities of 10-100 mJy kpc$^2$, see 
Fig. 1) will in principle be feasible with at least a similar 
signal-to-noise ratio across the whole Galaxy. In the ``early science'' 
phase of SKA1 the detection of jets from brown dwarfs will reach 300 pc 
and in the case of massive protostars, it will cover most of the Galaxy.

Observations at different frequencies across band 5 (providing angular 
resolutions in the 30 to 80 mas range) will determine the variations of 
the physical parameters along the jet axis and to image the region 
around the injection radius of the ionized gas in the jet. However, the 
high angular resolution study of radio jets with SKA1-MID only will be 
possible for jets of relatively high flux density (a few tenths of mJy). 
For weaker jets the full SKA (SKA2) will be required, since the uniform 
weighting SKA1 sensitivity is expected to be several times worst than 
for natural weighting (unless the antenna configuration is changed to 
favor longer baselines). For example, mapping the jet of a proto-brown 
dwarf is expected to take about 1.5 hours with SKA2, while it would take 
several hundreds of hours with SKA1.

A search for linear polarization using SKA1-MID will be feasible in the 
jets with relatively bright non-thermal knots. These knots have 
characteristic flux densities of about 100 $\mu$Jy. Assuming a linear 
polarization degree of 10\%, the detection of the Stokes parameters at a 
signal-to-noise ratio of 10, will require on-source integration times of 
order 30 minutes (about 2 hours in the ``early science'' phase of SKA1). 
With the full SKA (SKA2) it will be possible to search for linear 
polarization, and even to perform a high angular resolution mapping, in 
a much more extended sample.

For radio jets with continuum flux densities of order 3 mJy, a peak RRL 
flux density of 75 $\mu$Jy is expected (assuming a line width of about 
100 km s$^{-1}$, see formula 4.1).  Adopting a velocity resolution 
comparable to the line width, a 6-$\sigma$ detection will be achieved 
with an on-source integration time of 1 hour with SKA1-MID (4 h in the 
``early science'' phase of SKA1). If the emission is resolved spatially 
or we want better velocity resolution, the on-source time will increase 
accordingly. On the other hand, many recombination lines will be 
available in the SKA bands and it will be possible to stack them and 
improve the detectability (there are 35 RRLs in band 5, improving 
signal-to-noise ratio by a factor of $\sim$6). This type of observations 
will reach its full potential with the full SKA (SKA2), when it will be 
possible to map the radial velocity distribution of the jet and in 
combination with proper motions obtain the 3D kinematics.

Finally, the high sensitivity and angular resolution of SKA will make it 
possible to study the equatorial winds, probably driven by radiation 
pressure acting on the gas on the surface layers of the disk, that have 
been observed in a number of massive young stars, and to distinguish 
this kind of sources from jets, as has been done in S140 IRS1 with 
MERLIN by Hoare (2006).

\acknowledgments GA acknowledges support from MICINN (Spain) grant 
AYA2011-30228-C03 (co-funded with FEDER funds) and partial support from 
Junta de Andaluc\'{\i}a (TIC-126). LFR and CC-G acknowledge support from 
DGAPA, UNAM and CONACyT (Mexico).

\end{document}